\pdfoutput=1
\documentclass{vgtc}                          
\ifpdf
  \pdfoutput=1\relax                   
  \usepackage{graphicx}                
  \DeclareGraphicsExtensions{.pdf,.png,.jpg,.jpeg} 
\else
  \usepackage{graphicx}                
  \DeclareGraphicsExtensions{.eps}     
\fi%

\usepackage{microtype}                 
\PassOptionsToPackage{warn}{textcomp}  
\usepackage{textcomp}                  
\usepackage{mathptmx}                  
\usepackage{times}                     
\usepackage{cite}                      
\usepackage{tabu}                      
\usepackage{booktabs}                  
\usepackage{graphicx}
\graphicspath{ {figures/} }

\newcommand{\toolName}[0]{\textit{Visilence}}

\onlineid{1023}

\vgtccategory{Research}

\vgtcinsertpkg

\title{\toolName{}: An Interactive Visualization Tool for Error Resilience Analysis}

\author{Shaolun Ruan\thanks{e-mail: haywardryan@foxmail.com}\\ %
        \scriptsize Kent State University %
\and Yong Wang\thanks{e-mail: yongwang@smu.edu.sg }\\ %
     \scriptsize Singapore Management University %
\and Qiang Guan\thanks{e-mail: qguan@kent.edu}\\ %
     \scriptsize Kent State University %
}

\teaser{
  \centering
  \includegraphics[width=0.95\linewidth,height=100mm]{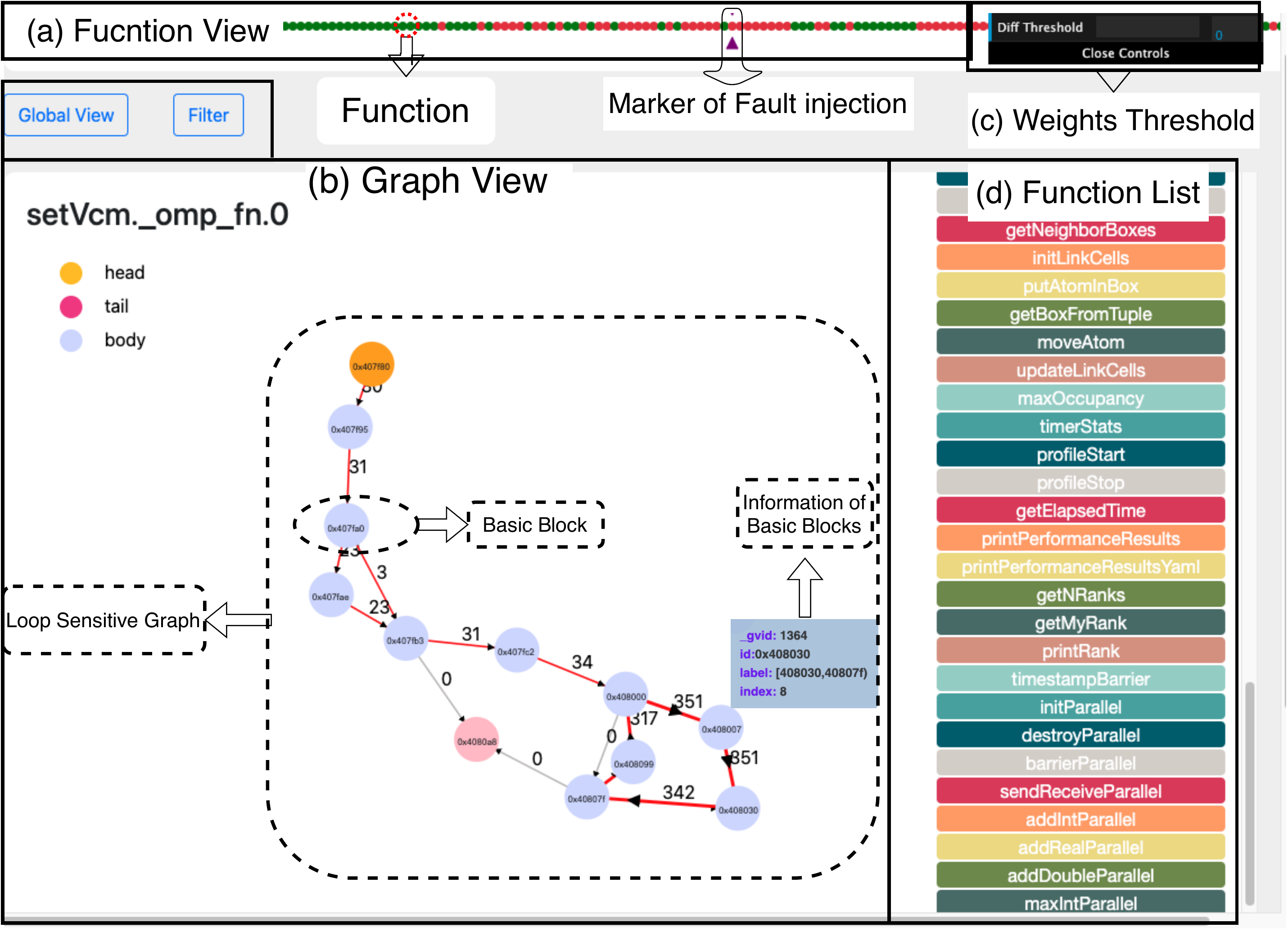}
  \caption{
  The interface of Visualization Engine. (a) Function view is a series of dots at top represent the functions. (b) The graph is shown in the Graph view and the nodes are basic blocks. (c) Weight threshold is used to set the weight threshold (d) The functions with specific names are listed in Function List. 
  }
  \label{fig:interface}
}

\abstract{
Soft errors have become one of the major concerns for HPC applications, as those errors can result in seriously corrupted outcomes, such as silent data corruptions (SDCs). Prior studies on error resilience have studied the robustness of HPC applications. However, it is still difficult for program developers to identify potential vulnerability to soft errors.
In this paper, we present \toolName{}, a novel visualization tool to visually analyze error vulnerability based on the control-flow graph generated from HPC applications. \toolName\ efficiently visualizes the affected program states under injected errors and presents the visual analysis of the most vulnerable parts of an application. We demonstrate the effectiveness of \toolName{} through a case study.
}

\CCScatlist{
  \CCScatTwelve{Human-centered computing}{Visu\-al\-iza\-tion}{Visu\-al\-iza\-tion techniques}{Treemaps};
  \CCScatTwelve{Human-centered computing}{Visu\-al\-iza\-tion}{Visualization design and evaluation methods}{}
}

\begin{document}

\firstsection{Introduction}

\maketitle

As the HPC systems keep scaling up, the chance of the systems encountering soft errors also increases~\cite{snir2014addressing}. Though many soft errors can be detected and corrected by hardware-level mechanisms, some errors escape these mechanisms and further propagate to the application-level~\cite{cho2013quantitative}, which can lead to a failure of applications and even serious outcomes such as silent data corruptions (SDCs). 
Prior studies~\cite{sridharan2009eliminating,wibowo2017characterizing} 
have shown that the impact of errors is not uniformly distributed, and the likelihood of causing SDC is also different. Soft errors affect different states of an application. It indicates that understanding how an error occurring on a particular program state affects the outcome of an application, which may help developers add extra protection in development, e.g., duplication of variables or instructions.

However, resilience analysis for HPC applications is often known as a ``Black Box'' analysis: the user can estimate the resilience characteristics via fault injection~\cite{FSEFI} to an application, which usually lacks explainability on a case-by-case basis.
For example, SpotSDC~\cite{li2020spotsdc} visualizes error propagation in programs through fault injection methods, whereas the targeted vulnerable code regions and functions in SpotSDC require the user's expertise, which is not appropriate for general developers. 
The conventional preceding studies rarely analyze error propagation and resilience through visualization methods and symbols turning inflexible HPC programs trace data into graphical representations and providing interactive analysis modes.

We propose a novel control-flow based visualization tool to explore the error resilience of HPC applications. Furthermore, we showcase the error propagation pattern along with the basic blocks of an example faulty run of CoMD and demonstrate the usage of \toolName\ to identify the critical sections of the applications.

\section{\toolName}

We roughly categorize the introduction of \toolName\ into two categories: overall workflow of \toolName{} and generation pipeline of visualization.

\subsection{Overall workflow of \toolName{}}

At a high level, \toolName\ needs three levels of abstractions: (a) a model that can keep the static and dynamic program states, (b) a format to allow systematic analysis of the program states, and (c) a visualization tool that offers a friendly interface to identify the code regions that are sensitive to the errors for the users. We define Loop Sensitive graph (LSG) generated from the dynamic traces and Critical Vector Graph (CVG) generated based on the accumulation of multiple LSGs. The workflow of \toolName\ proceeds as follows: \textit{(i)}, it takes an HPC program as input and conducts a statistic fault injection campaign on the application to generate a set of dynamic execution traces; \textit{(ii)}, it creates LSGs/CVGs based on the obtained dynamic traces of the application, and \textit{(iii)} it implements a novel visualization system that takes the LSGs/CVGs as the data source and provides a fine-grained representation of error propagation and resilience characteristic for the application.

\subsection{Generation Pipeline of Visualization}

\begin{figure}[tbhp]

\centering
\includegraphics[width=0.8\columnwidth]{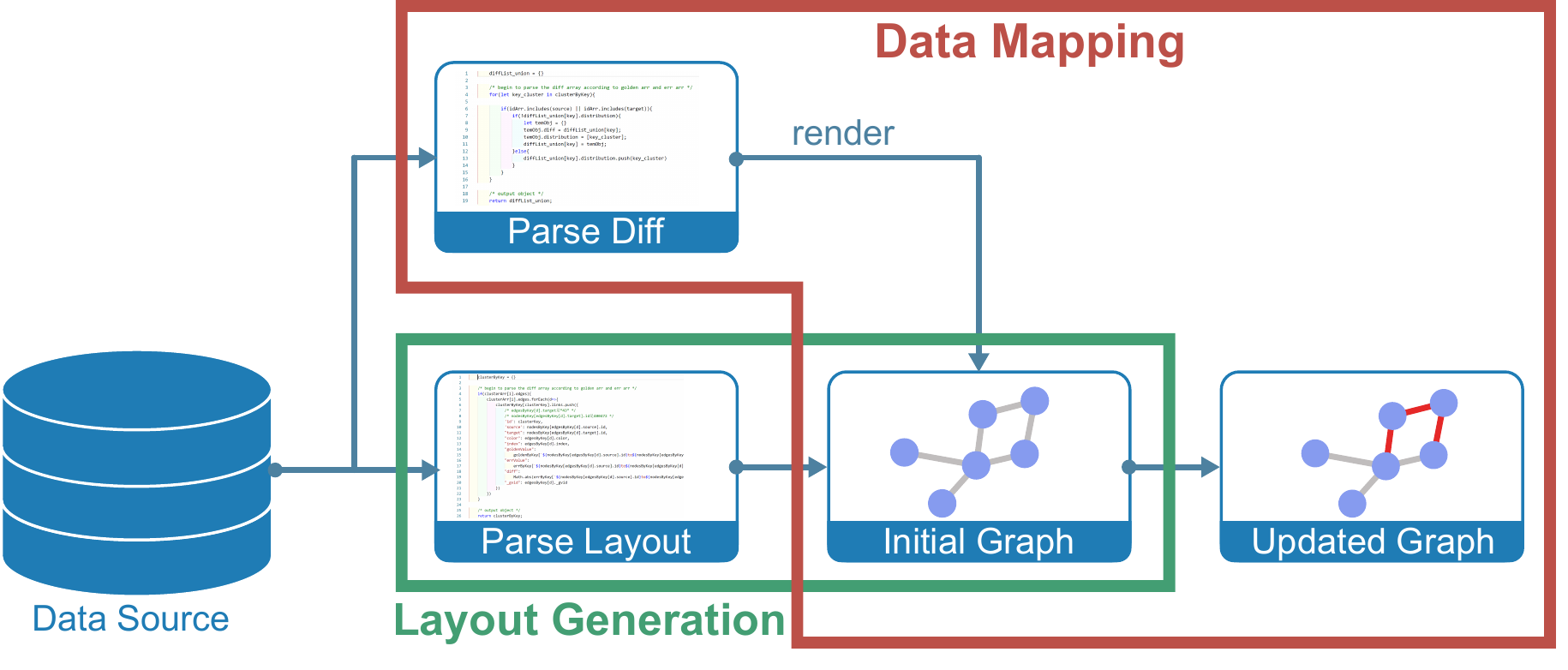}
\caption{The workflow of our visualization system}
\label{fig_workflow}

\end{figure}

\toolName{}
has two modules, namely the function selecting module and the graph module, to support the collaborative design of basic-block like visualization. The pipeline is shown in Fig. \ref{fig_workflow}. The visualization system has two separated stages for resilience graph generation, namely the layout simulation and the anomaly mapping. 

We implement a user-friendly interface to visualize error propagation and functions interactively (see Fig. \ref{fig:interface}). The interface consists of four parts:

\begin{itemize}

    \item \textbf{Function View (a)} is a sequence of functions which are represented by dots. These functions are placed in the order of where they are defined. A green dot means it matches exactly like the golden run`s, or it would be rendered in red when they are different in weights. The triangle on the sequence is a marker labeling the function where the fault is injected. 
    
    \item \textbf{Graph View (b)} shows the Loop Sensitive Graph/Critical Vector graph. The vertices of the graph are basic blocks and the \textit{head} (in yellow) and \textit{tail} (in red) nodes are the entry and exit of the function respectively. The edges represent the connections between two basic blocks in the CFG, and the weights are the absolute values of the different executed times between the faulty traces and golden runs. The edge is gray when its weight is zero and is red otherwise. There are two options above: Global view and Filter. 
    \item \textbf{Weight Threshold (c)} is used to filter the edges. When we slide the bar in Weight Threshold, the value would be adjusted, and the edges with smaller weights below the threshold would be assigned into gray. 
    
    \item \textbf{Function List (d)} lists all the functions in the program with specific name in the same order in \textbf{Function View}. We can click on it to select the function to be shown in \textbf{Graph View}.
    
\end{itemize}



\section{Case study}

When soft errors occur in the running process of the program, this error may affect the subsequent control flow. Our tool can intuitively indicate how this error propagates.

Fig.~\ref{fig:interface} shows the error propagation pattern along with the basic blocks of an example faulty run of CoMD~\cite{comd}. The series of dots at the top represents all the 157 functions of CoMD. The green dot indicates that the LSG generated for that function is consistent with the golden run, while the red dot indicates that they are inconsistent. The ``marker of fault injection" indicates that the fault was injected in that function.

Fig.~\ref{fig:interface} presents an example of LSG for the function 'setVcm\_omp\_fn.o' in benchmark program CoMD. The function starts from the `head' basic block $'0x407f80'$ and ends in the `tail' basic block $0x4080a8$, in total 12 basic blocks. The weights are the difference in executed times between the golden run and the faulty run. The biggest difference in this function is 351 on the edges from basic block $0x408000$ to $0x408030$. The path from the basic block $'0x408000'$ to $'0x408030'$ maps to the source code of `initAtoms.c' at Lines 126 to 129 inside a for loop. We observed that 64 functions were affected by the injected fault.

\section{conclusion}

We proposed \toolName{}, a control-flow graph based visualization tool for error resilience analysis, which provides human analysts with detailed facets of error propagation for further decision making.
\toolName{} addresses the issue of understanding how the applications are affected by the errors via a graph-based abstraction to represent the affected program states and the reason for the error propagation across different error scenarios.

\bibliographystyle{abbrv-doi}

\bibliography{template}
\end{document}